\documentclass[
reprint,
superscriptaddress,
amsmath,amssymb,
aps,
prl,
]{revtex4-2}

\usepackage{graphicx}
\usepackage{dcolumn}
\usepackage{bm}
\usepackage{gensymb}
\usepackage{url}
\usepackage{multirow}

\usepackage{xcolor}
\usepackage{orcidlink}

\begin{document}

\title{ Synthetic crystal rotation with spacetime metamaterials }

\author{Iñigo Liberal \orcidlink{0000-0003-1798-8513}}
\thanks{Corresponding author: inigo.liberal@unavarra.es}
\affiliation{%
 Department of Electrical, Electronic and Communications Engineering, Institute of Smart Cities (ISC), Public University of Navarre (UPNA), 31006 Pamplona, Spain
}%

\author{Alejandro Manjavacas \orcidlink{0000-0002-2379-1242}}
\affiliation{%
 Instituto de Qu\'imica F\'isica Blas Cabrera (IQF), CSIC, 28006 Madrid, Spain
}%

\begin{abstract}

The interaction of light with rotating bodies has been historically limited to rotation frequencies much smaller than optical frequencies. Here, we investigate synthetic crystal rotations, i.e., spatiotemporal modulations mimicking the rotation of an anisotropic crystal, which grant access to large rotation frequencies. Spatiotemporal modulations change the fundamental symmetries of the electromagnetic field, breaking temporal and rotation symmetries, but preserving a spatiotemporal rotation symmetry that enforces the conservation of a combination of energy and spin angular momentum (SAM). The scattering of optical pulses by synthetically rotating crystals results in spatiotemporal light with intra-pulse SAM changes. The frequency-domain response reveals sidebands with frequency/SAM locking, and negative frequency sideband transitions for large enough rotation frequencies. Our results highlight the qualitatively different light-matter interaction regimes accessed by synthetic rotations. 

\end{abstract}

\maketitle

Rotation is a ubiquitous and fundamental concept in physics. It represents one of the most basic forms of motion \cite{Tipler2007physics}, embodies one of nature's fundamental symmetries \cite{Schwichtenberg2018physics}, and has profound connections with general relativity \cite{Frauendiener2018notes,Kish2022quantum}. Following Sagnac's effect \cite{Post1967sagnac,Anderson1994sagnac}, the interaction of light with mechanical rotations has been the focus of multidisciplinary research. Examples of rotation-induced optical phenomena include the rotational Doppler effect \cite{Emile2023rotational} and its nonlinear extensions \cite{Li2016rotational}, stimulated light emission and inelastic scattering \cite{Asenjo2011stimulated}, bianisotropy and nonreciprocity \cite{Shiozawa1973phenomenological,Zel1986rotating}, distinct diffraction and interference features \cite{Mazor2019rest,De1980scattering}, and wave instabilities \cite{Lannebere2016wave}. In the context of fluctuation-induced phenomena, rotation is at the origin of  vacuum rotational friction \cite{Manjavacas2010vacuum,Manjavacas2010thermal,Sanders2019nanoscale,Deop2023control} and gives rise to lateral Casimir forces \cite{Manjavacas2017lateral}. Rotation also generates \cite{Torovs2022generation} and modifies \cite{Cromb2023mechanical} entanglement, facilitating the quantum sensing of Earth's rotation \cite{Silvestri2024experimental}. Beyond fundamental scientific motivations, rotating optical systems have technological applications in laser scanning \cite{Li1995laser,Wang2020mems} and gyroscopes \cite{Lefevre2022fiber,dell2023miniaturization}.

Nevertheless, the operation of optical rotating systems is limited to the regime of rotation frequencies much smaller than the frequency of light, $\Omega\ll\omega_0$, due to the difficulty of accelerating matter to high speeds. Despite tremendous technological advances, even state-of-the-art experiments with levitated nanoparticles have rotation frequencies limited to the GHz range \cite{Ahn2018optically,Reimann2018ghz}. 

Recently, there has been a renewed interest in the ultra-fast modulation of material parameters leading to the field of time-varying media \cite{Galiffi2022photonics}, 4D optics \cite{Engheta2023four} and/or spacetime metamaterials \cite{Caloz2019spacetime}. Within this context, synthetic motion  refers to spatiotemporal modulations that mimic the response of moving matter \cite{Caloz2022generalized,Bahrami2023electrodynamics,Mazor2019one,Harwood2024super}. Importantly, because no matter is physically being moved, synthetic motion allows for high speeds, and even the access to luminal and superluminal regimes forbidden for moving matter \cite{Harwood2024super,Deck2018wave,Pendry2022photon,Galiffi2019broadband}. In addition, spatiotemporal modulations have been shown to alter the fundamental symmetries of the electromagnetic field, leading to reduced spatiotemporal symmetries and modified conservation laws \cite{Liberal2024spatiotemporal,Ortega2023tutorial,Jajin2024symmetries}. However, the study of rotation-like modulations has been largely restricted to perturbative modulations empowering circulators \cite{Sounas2017non,Pakniyat2024magnet,Serra2023rotating}, nonreciprocal amplifiers \cite{Galiffi2022} and microwave metasurfaces for orbital angular momentum (OAM) control \cite{Moussa2022penrose}.

\begin{figure}[!t]
  \centering
    \includegraphics[width=0.99\columnwidth]{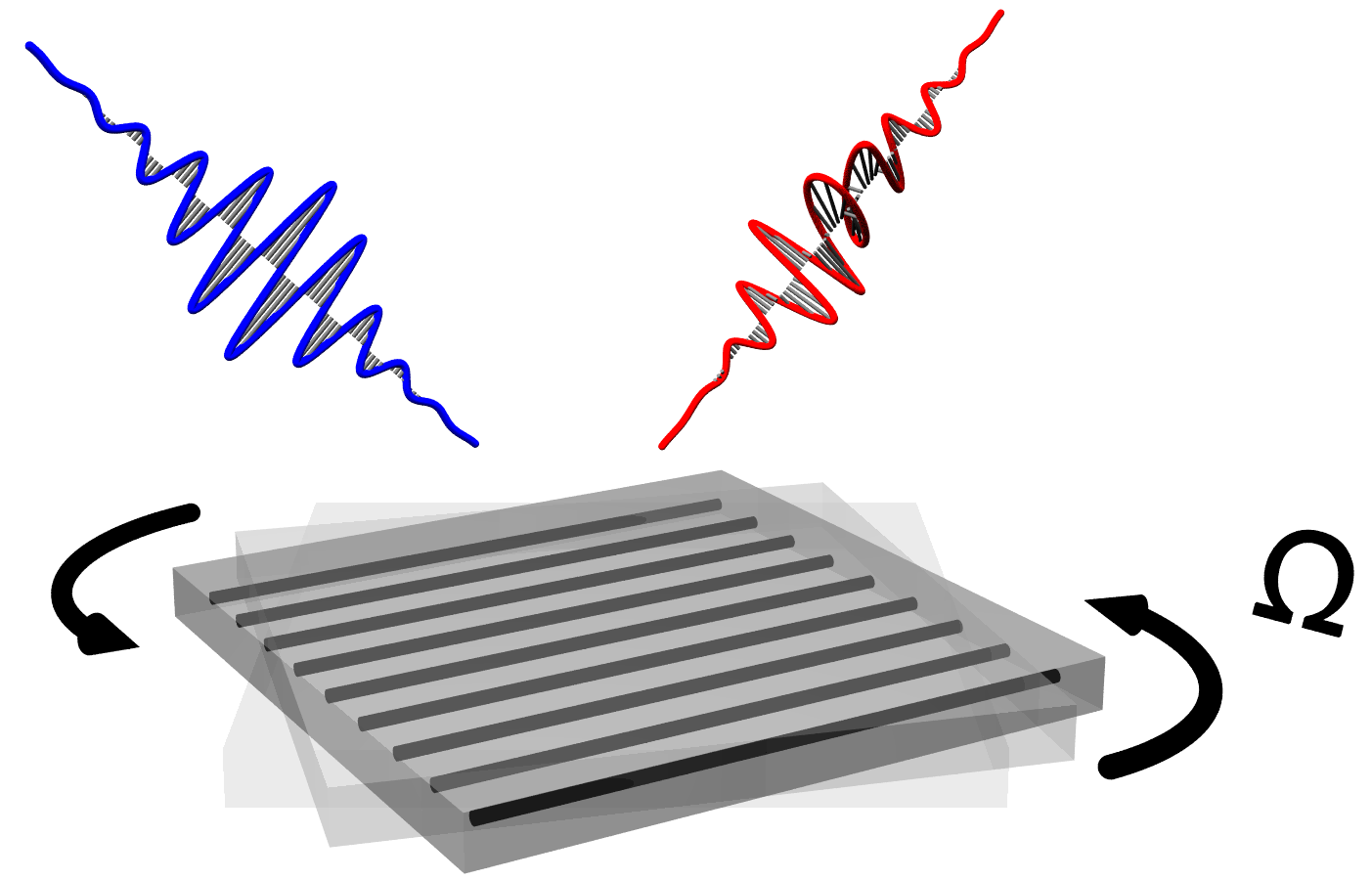}
  \caption{{\bf Concept.} Illustration of synthetic crystal rotation implemented via spatiotemporal modulations that emulate the rotation of an anisotropic crystal. }
  \label{fig:Schematic}
\end{figure}

Here, we investigate synthetic crystal rotation, i.e., spatiotemporal modulations that mimic the rotation of an anisotropic crystal (see Fig.\,\ref{fig:Schematic}). Specifically, we clarify their fundamental spatiotemporal symmetries and conservation laws, explore how high rotation frequencies provide access to qualitatively different phenomena both in the transient and frequency domains, and discuss practical implementation strategies at optical frequencies.
   
As schematically depicted in Fig.\,\ref{fig:Schematic}, we consider a material layer with a spatiotemporal permittivity modulation mimicking the rotation of an anisotropic crystal around the z-axis:
\begin{equation}
\boldsymbol{\varepsilon}\left(z,t\right)=\mathbf{R}\left(\Omega t\right)\cdot\boldsymbol{\varepsilon}_{S}\left(z\right)\cdot\mathbf{R}^{-1}\left(\Omega t\right),
\label{eq:eps_S}
\end{equation}
where 
$\boldsymbol{\varepsilon}_{S}\left(z\right)=\widehat{\mathbf{x}}\widehat{\mathbf{x}}\,\varepsilon_x(z)+\widehat{\mathbf{y}}\widehat{\mathbf{y}}\,\varepsilon_y(z)$ is a static (no time-modulated) anisotropic permittivity, and 
$\mathbf{R}\left(\phi\right)={\rm cos}\phi(\widehat{\mathbf{x}}\widehat{\mathbf{x}}+\widehat{\mathbf{y}}\widehat{\mathbf{y}})
+{\rm sin}\phi(\widehat{\mathbf{y}}\widehat{\mathbf{x}}-\widehat{\mathbf{x}}\widehat{\mathbf{y}})
$ is the rotation dyadic around the z-axis. We assume that the system is excited by a plane-wave field propagating along the z-axis. As a result, all fields are transversal and the vector potential reduces to $\mathbf{A}\left(\mathbf{r},t\right)=\widehat{\mathbf{x}}A_{x}\left(z,t\right)+\widehat{\mathbf{y}}A_{y}\left(z,t\right)$, defining the electric $\mathbf{E}\left(z,t\right)=-\partial_t\mathbf{A}\left(z,t\right)$ and magnetic $\mathbf{B}\left(z,t\right)=\nabla\times\mathbf{A}\left(z,t\right)$ fields. The dynamical behaviour of the fields is determined by Maxwell's equations, which are equivalent to the following wave equation for the vector potential
\begin{equation}
\partial_{z}^{2}\mathbf{A}\left(z,t\right)=c^{-2}\partial_{t}\left\{ \boldsymbol{\varepsilon}\left(z,t\right)\cdot\partial_{t}\mathbf{A}\left(z,t\right)\right\}.
\label{eq:waveeq}
\end{equation}

{\it Spatiotemporal symmetries and conserved quantities.--} Spatiotemporal modulation alters the fundamental symmetries of the system. In particular, the modulation described by Eq.~(\ref{eq:eps_S}) manifestly breaks both time-translation and rotational symmetries. At the same time, the system retains a spatiotemporal rotation symmetry, corresponding to a  simultaneous time translation by $dt$ and rotation of the optical axis by an angle $\Omega dt$. Mathematically, $\boldsymbol{\varepsilon}\left(z,t\right)$ satisfies the following property
\begin{equation}
\partial_{t}\boldsymbol{\varepsilon}\left(z,t\right)=\Omega\,\left[d\mathbf{R},\boldsymbol{\varepsilon}\left(z,t\right)\right],
\label{eq:eps_sym}
\end{equation}
where $d\mathbf{R}=\widehat{\mathbf{y}}\widehat{\mathbf{x}}-\widehat{\mathbf{x}}\widehat{\mathbf{y}}$ is the differential rotation dyadic. 

Noether's theorem establishes that any continuous symmetry is associated with a conserved quantity \cite{Cohen1998atom}. To exploit this connection, we describe the system using the Lagrangian density 
$\mathcal{L}\left(z,t\right)=\frac{\varepsilon_{0}}{2}\,\partial_{t}\mathbf{A}\left(z,t\right)\cdot\boldsymbol{\varepsilon}\left(z,t\right)\cdot\partial_{t}\mathbf{A}\left(z,t\right)-\frac{\varepsilon_{0}c^{2}}{2}\,\left(\partial_{z}\mathbf{A}\left(z,t\right)\right)^{2}$, which is justified by the fact that the Euler-Lagrange equation recovers the correct wave equation for the vector potential. According to Noether’s theorem, if the system is invariant under a continuous transformation of the dynamical variables $\mathbf{A}'=\mathbf{A}+d\mathbf{A}$, then the following quantity is conserved \cite{Ortega2023tutorial}
\begin{equation}
\psi\left(d\mathbf{A}\right)=\int dz\,\left\{ \frac{\partial\mathcal{L}}{\partial\partial_{t}\mathbf{A}}\cdot d\mathbf{A}-\mathcal{L}\,dt\right\}.
\label{eq:Noether}
\end{equation}

Analyzing Eq.~(\ref{eq:eps_sym}), we define the following spatiotemporal rotation symmetry
\begin{equation}
d\mathbf{A}\left(z,t\right)=dt\,\left\{\partial_{t}\mathbf{A}\left(z,t\right)-\Omega \,\,d\mathbf{R}\cdot\mathbf{A}\left(z,t\right)\right\},
\label{eq:dA}
\end{equation}
which, as shown in \cite{SM}, preserves the form of the wave equation given in Eq.~(\ref{eq:waveeq}). Substituting Eq.~(\ref{eq:dA}) into Eq.~(\ref{eq:Noether}), we find that the conserved quantity associated with spatiotemporal rotations is a linear superposition of energy and spin angular momentum (SAM), weighted by the rotation frequency $\Omega$
\begin{equation}
\psi\left(d\mathbf{A}\right)=H-\Omega J_{S,z}. \nonumber
\end{equation}
Here, $H=\int dz\,\,h(z,t)$ represents the total energy of the system and $h(z,t)=\frac{\varepsilon_{0}}{2}\,\partial_{t}\mathbf{A}\left(z,t\right)\cdot\boldsymbol{\varepsilon}\left(z,t\right)\cdot\partial_{t}\mathbf{A}\left(z,t\right)+\frac{\varepsilon_{0}c^{2}}{2}\,\left(\partial_{z}\mathbf{A}\left(z,t\right)\right)^{2}$ is the corresponding energy density. Similarly, $J_{S,z}=\int dz\,j_{S,z}$ is the total SAM with $j_{S,z}=-\widehat{\mathbf{z}}\cdot (\varepsilon_{0}\boldsymbol{\varepsilon}\left(z,t\right)\cdot\partial_{t}\mathbf{A}\left(z,t\right)\times\mathbf{A}\left(z,t\right))$ being its correspoding density. As anticipated, since both time translation and rotational symmetries are broken, neither energy nor SAM is conserved. However, these symmetries combine into a single spatiotemporal rotational symmetry, which results in the conservation of a linear combination of energy and SAM. This conservation law provides the framework for light-matter interactions with synthetically rotating crystals.

{\it Continuity equations.--} The conservation of $H-\Omega J_{S,z}$ can alternatively be proven by explicitly evaluating the time derivatives of energy and SAM. This approach allows the construction of the following continuity equations, which provide important insights into their fluxes and sources/sinks \cite{SM}
\begin{align}
\partial_{t}h+\partial_{z}F_{h}
&=-\frac{\varepsilon_{0}}{2}\,\partial_{t}\mathbf{A}\cdot\partial_{t}\boldsymbol{\varepsilon}\cdot\partial_{t}\mathbf{A},
\label{eq:Continuity_h}\\
\partial_{t}j_{S,z}+\partial_{z}F_{S}
&=-\frac{\varepsilon_{0}}{2}\partial_{t}\mathbf{A}\cdot\left[d\mathbf{R},\boldsymbol{\varepsilon}\right]\cdot\partial_{t}\mathbf{A},
\label{eq:Continuity_SAM}
\end{align}
where the flux of energy is the Poynting vector $F_{h}\left(z,t\right)=-\varepsilon_{0}c^{2}\,\partial_{t}\mathbf{A}\left(z,t\right)\cdot\partial_{z}\mathbf{A}\left(z,t\right)$, while the flux of SAM is 
$F_{S}\left(z,t\right)=\varepsilon_{0}c^{2}\mathbf{A}\left(z,t\right)\cdot\mathbf{B}\left(z,t\right)$. The presence of source/sink terms on the right hand side of  Eqs.~(\ref{eq:Continuity_h}) and (\ref{eq:Continuity_SAM}) confirms that neither the energy nor the SAM are conserved quantities. At the same time, it is evident that the linear combination of these sources, weighted by  $\Omega$, cancels out, thereby confirming the conservation of of $H-\Omega J_{S,z}$.

The critical advantage of synthetic rotations is that, since no matter is physically being moved, they provide access to rotation frequencies comparable or even greater that the frequency of light. As a result, such high rotation frequencies enable the access to qualitatively different regimes for light-matter interactions both in transient and frequency-domain configurations.

{\it Transient response.--} To illustrate the proposed concept of synthetic crystal rotation, we begin by investigating the transmission/reflection from a thin material layer of thickness $d$, whose permittivity is given by Eq.~(\ref{eq:eps_S}). The film is excited by a transversal plane-wave pulse with vector potential $\mathcal{\boldsymbol{A}}_{0}\left(z,t\right)=\widehat{\mathbf{x}}\,A_{0}{\rm sech}[(z-z_0-ct)/\lambda_0]\,{\rm sin}[2\pi/\lambda_0(z-ct)]$. In the thin-film limit, the response of the layer can be approximated by a surface polarization distribution $\mathcal{\boldsymbol{P}}\left(z,t\right)=\mathbf{p}\left(t\right)\,\delta\left(z\right)=\boldsymbol{\alpha}\left(t\right)\cdot\mathcal{\boldsymbol{E}}\left(z,t\right)\delta\left(z\right)$. The total field is given by the sum of the incident and the fields radiated by the induced polarization, $\mathcal{\boldsymbol{E}}\left(z,t\right)=\mathcal{\boldsymbol{E}}_{0}\left(z,t\right)+\mathcal{\boldsymbol{E}}_{P}\left(z,t\right)$, whose value can be found by solving a boundary value equation on the thin-film $\mathcal{\boldsymbol{E}}\left(0,t\right)=\boldsymbol{\alpha}^{-1}\left(t\right)\mathbf{p}\left(t\right)=\mathcal{\boldsymbol{E}}_{0}\left(0,t\right)-\frac{Z_{0}}{2}\,\partial_{t}\mathbf{p}\left(t\right)$ (see \cite{SM} for details). 

\begin{figure}[!t]
  \centering
    \includegraphics[width=0.9\columnwidth]{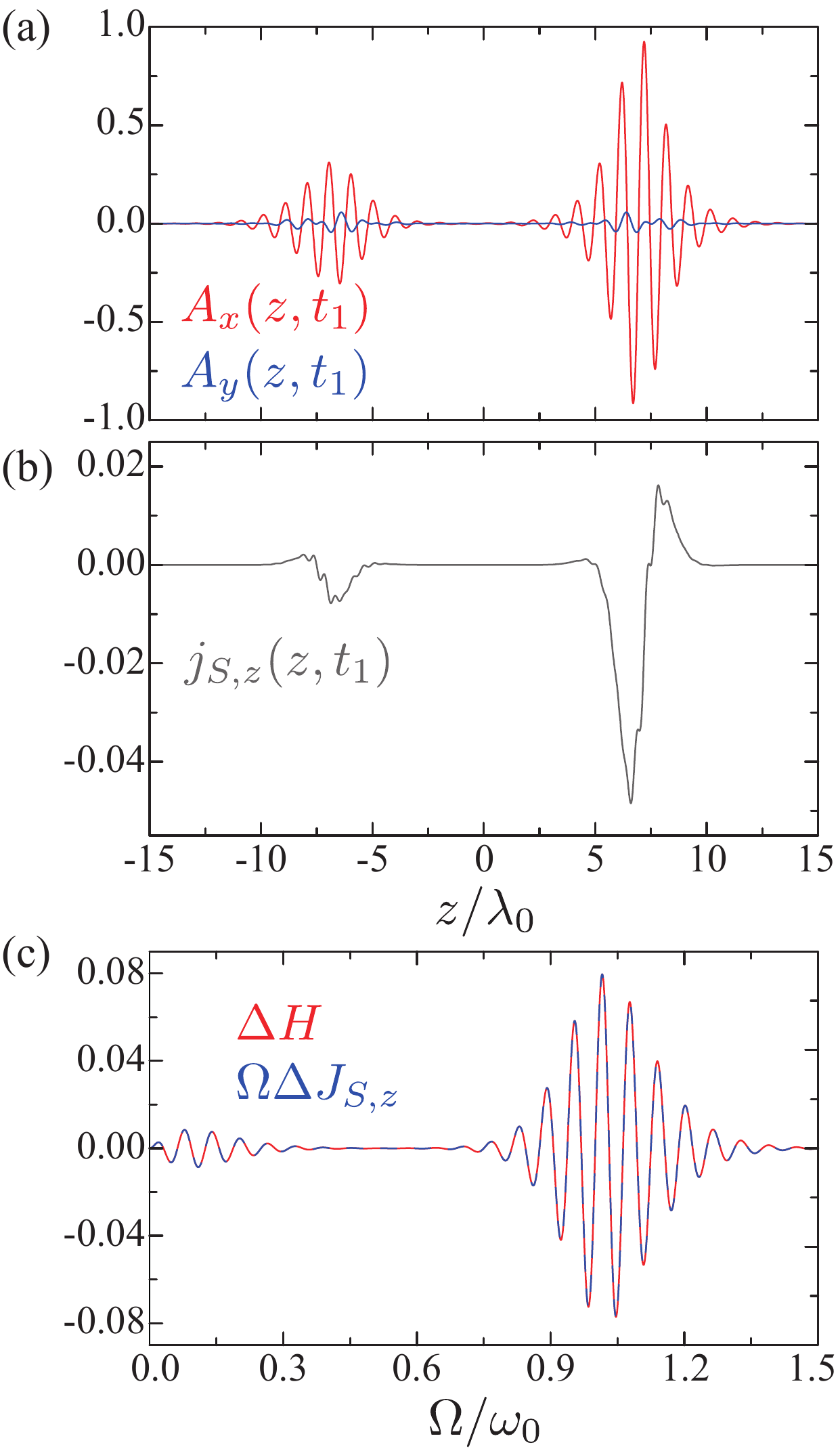}
  \caption{{\bf Transient response.} Snapshots of the vector potential $\mathbf{A}$ (a) and  the SAM density $j_{S,z}$ (b) at $t_1=15 T_0$, for $\Omega=\omega_0/10$, $z_0=-8\lambda_0$, $\varepsilon_x=2.5$, $\varepsilon_y=2$, $d=\lambda_0/10$. Notice that $j_{S,z}$ is normalized to $\omega_0\varepsilon_0 A_0^2$. (c) Variation in the total energy $\Delta H=H(t)-H(0)$ and SAM $\Delta J_{S,z}=J_{S,z}(t)-J_{S,z}(0)$, as a function of $\Omega$, normalized to the initial energy $H(0)$.}
  \label{fig:Ultrafast}
\end{figure}

Figure~\ref{fig:Ultrafast}(a) shows a snapshot of the resulting vector potential after the interaction with the thin-film. A complete video animation and additional examples can be found in \cite{SM}. Intuitively, the field consists of reflected and transmitted pulses, containing cross-polarized components resulting from the anisotropy introduced by the rotation. Remarkably, because the rotation frequency is not necessarily small ($\Omega=\omega_0/10$ in this example), the phase-shift between the co- and cross-polarized fields changes within the pulses. This effect is more clearly illustrated in Fig.~\ref{fig:Ultrafast}(b), which presents a snapshot of the SAM density. The results shown in this panel confirm that both transmitted and reflected pulses carry nonzero SAM, with the SAM density changing sign within the pulses. In other words, the pulses scattered from synthetically rotating crystals consist of nonseparable polarization-pulse shape solutions to Maxwell's equations, commonly referred to as spatiotemporal light fields \cite{Shen2023roadmap,Zhan2024spatiotemporal}. The characteristics of the pulse shape and SAM content can be controlled by engineering the incident pulse shape and rotation frequency. A few examples are reported in \cite{SM}. 

The variations in the total energy $\Delta H=H(t)-H(0)$ and SAM $\Delta J_{S,z}=J_{S,z}(t)-J_{S,z}(0)$, after the interaction with the thin-film, are analyzed in Fig.~\ref{fig:Ultrafast}(c) as a function of $\Omega$. Due to the spatiotemporal symmetry of the system, these variations are perfectly correlated, such that $\Delta H=\Omega\Delta J_{S,z}$. In practice, this implies that amplification of the field and the generation of SAM are locked. Other important features of $\Delta H$ and $\Delta J_{S,z}$ inferred from these results include the presence of fast oscillations related to the synchronization of the pulse peaks with the optical axis, as well as slow changes in the net SAM content of the pulse, with additional examples reported in \cite{SM}. 

\begin{figure}[!t]
  \centering
    \includegraphics[width=0.95\columnwidth]{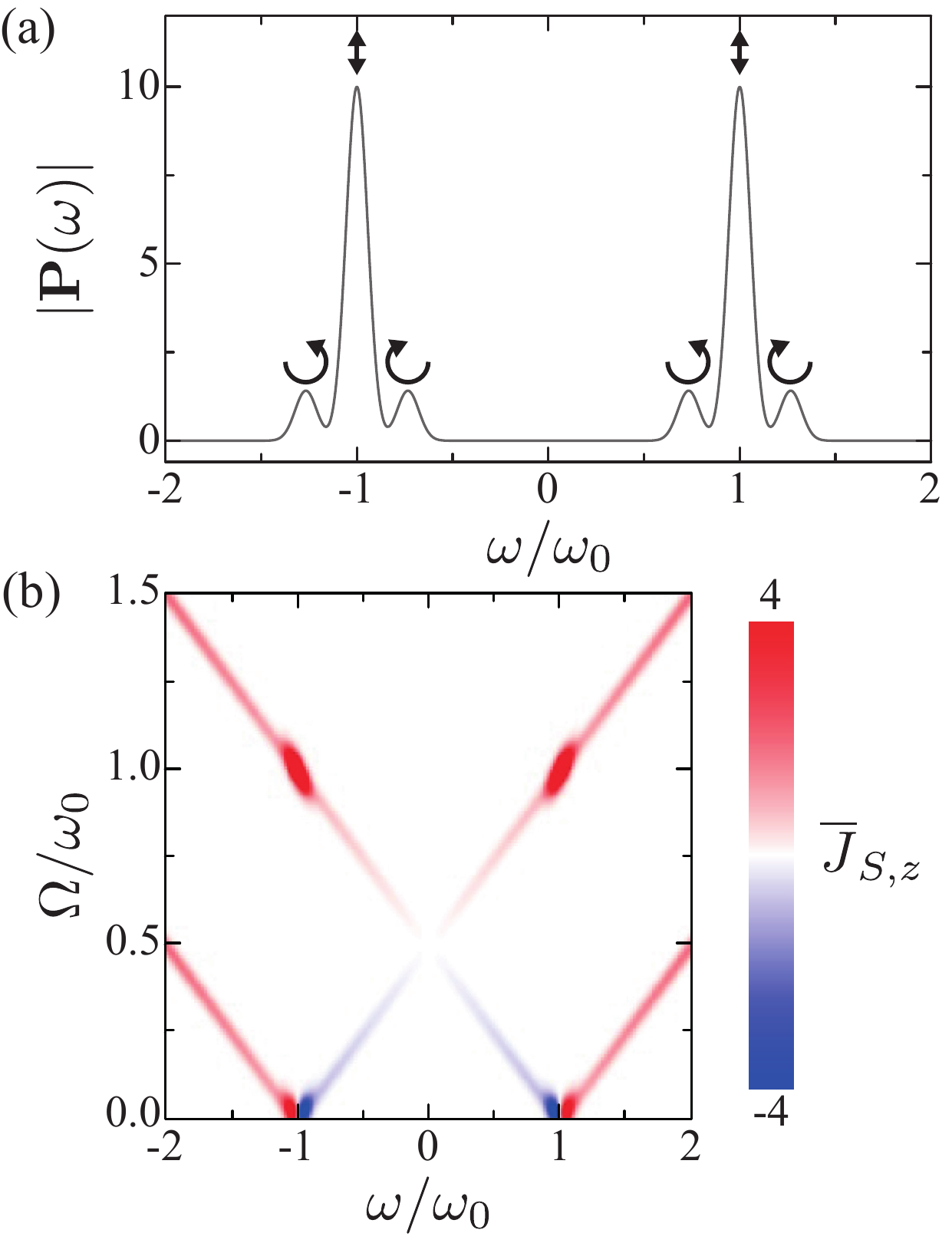}
  \caption{ {\bf Frequency-domain response}. (a) Induced polarization spectra for $\Omega=2\omega_0/15$ and $q=25$, normalized to $(\alpha_{x}-\alpha_{y})/8|\widetilde{E}(0)|$. (b) Time-averaged SAM of the reflected wave as a function of $\Omega$, normalized to $((\alpha_{x}-\alpha_{y})/16|\widetilde{E}(0)|)^2$.}
  \label{fig:Frequency}
\end{figure}

{\it Frequency-domain response.--} To complete our analysis, we consider the same thin-film excited by a narrow-band signal centered around frequency $\omega_0$, $\widetilde{\mathcal{E}}_{0}\left(0,\omega\right)=\widetilde{E}\left(\omega+\omega_{0}\right)+\widetilde{E}\left(\omega-\omega_{0}\right)$, with $\widetilde{E}\left(\omega\right)=E_{0}\,\exp[-\frac{q^{2}}{4}\frac{\omega^{2}}{\omega_{0}^{2}}]$ and $q$ being a dimensionless parameter that controls the width of the spectrum. The frequency-domain polarization induced on the thin-film can be obtained by constructing an iterative solution for the boundary value problem (see \cite{SM} for details). To first order, the induced polarization is given by 
\begin{align}
\widetilde{\mathbf{P}}&\left(\omega\right) =\widehat{\mathbf{x}}\,\frac{\alpha_{x}+\alpha_{y}}{4}\,\widetilde{\mathcal{E}}_{0}\left(0,\omega\right)
+\frac{\alpha_{x}-\alpha_{y}}{8}\times \nonumber \\  
&\left\{
\left(\widehat{\mathbf{x}}-i\widehat{\mathbf{y}}\right)\widetilde{\mathcal{E}}_{0}\left(0,\omega+2\Omega\right)+\left(\widehat{\mathbf{x}}+i\widehat{\mathbf{y}}\right)\widetilde{\mathcal{E}}_{0}\left(0,\omega-2\Omega\right)
\right\}. \nonumber
\end{align}
This equation shows that the frequency-domain response of the spatiotemporal thin film can be interpreted as a combination of (i) an effective linear and isotropic polarizability, $(\alpha_{x}+\alpha_{y})/4$, which produces linearly polarized radiation centered at $\omega_0$, and (ii) an effective nonlinear and anisotropic polarizability, $(\alpha_{x}-\alpha_{y})/8$, which generates two sidebands with opposite circular polarizations at $\omega_0\pm2\Omega$. 
In other words, the induced polarization exhibits frequency--SAM locking, schematically illustrated in Fig.~\ref{fig:Frequency}(a), which can be understood as a consequence of the conservation of $H-\Omega J_{S}$. Notably, this spectral response matches that predicted for rotating anisotropic particles \cite{Asenjo2011stimulated}, demonstrating that synthetic motion can replicate the effects of physically rotating matter. 

At the same time, synthetic motion offers the possibility of exploring new regimes made accessible by large rotation frequencies $\Omega$. To illustrate this possibility, Fig.~\ref{fig:Frequency}(b) depicts the spectrum of the resulting time-averaged SAM, $\mathbf{\overline{j}}_{S}=\frac{\varepsilon_{0}Z_{0}^{2}}{4}\int_{-\infty}^{\infty}d\omega\,i\omega\,\widetilde{\mathbf{P}}\left(\omega\right)\times\widetilde{\mathbf{P}}^{*}\left(\omega\right)$. For small rotation frequencies, $\Omega\ll\omega_0$, this spectrum is characterizeed by low- and high-frequency sidebands of opposing SAM sign, as previously discussed. However, the spectrum becomes qualitatively different as $\Omega$ becomes comparable to $\omega_0$, due to negative frequency transitions. 

First, at $\Omega=\omega_0/2$, the two low-frequency sidebands degenerate at $\omega=0$, yielding zero SAM, consistent with the fact that circular polarization cannot exist at zero frequency. At this point, the SAM spectrum becomes single-sideband. Second, for $\omega_0/2<\Omega<\omega_0$, the low-frequency sidebands cross from positive to negative frequencies, reversing their SAM sign in the process. In this regime, the SAM spectrum is characterized by asymmetrically shifted low- and high-frequency sidebands with the same sign. Next, at $\Omega=\omega_0$, the low-frequency sidebands interfere with the linearly polarized spectral line at $\omega=\omega_0$, resulting in a sudden increase in SAM. A similar effect is also observed when the rotation frequency is smaller than the bandwidth of the incident wave. Finally, for $\Omega>\omega_0$, the SAM spectrum consists of two high-frequency sidebands with the same sign. In summary, rotation frequencies on the order of the frequency of light provide access to qualitatively distinct spectral regimes. These qualitatively different regimes are expected to modify previously studied light-matter interactions with rotating bodies \cite{Emile2023rotational,Li2016rotational,Asenjo2011stimulated,Shiozawa1973phenomenological,Zel1986rotating,Mazor2019rest,De1980scattering,Lannebere2016wave,Manjavacas2010vacuum,Manjavacas2010thermal,Manjavacas2017lateral,Torovs2022generation,Cromb2023mechanical,Silvestri2024experimental}. 

{\it Practical implementations.--}
The physical implementation of synthetic crystal rotation naturally aligns with the optical Kerr effect in crystals, which offers a fast, polarization-sensitive nonlinear response. In particular, the polarization induced by a third-order optical nonlinearity is given by \cite{Boyd2008nonlinear}
$P_{i}=\chi^{(3)}_{ijkl}E_{j}E_{k}E_{l}$. For strong pump and weak signal fields, $E_{j}=E^{\rm P}_{j}+E_{j}^{\rm S}$, the optical Kerr effect leads to an effective susceptibility tensor of the form 
$\chi_{ij}^{\mathrm{eff}}=\left(\chi_{ijkl}^{(3)}+\chi_{ikjl}^{(3)}+\chi_{iklj}^{(3)}\right)E_{k}^{\mathrm{P}}E_{l}^{\mathrm{P}}$ (see \cite{SM} for details). For a circularly polarized pump, 
$\mathbf{E}^{\rm P}=E_{0}(\widehat{\mathbf{x}}{\rm cos}(\Omega t)+\widehat{\mathbf{y}}{\rm sin}(\Omega t))$,
and for specific crystal symmetries, the effective susceptibility tensor recovers the form of a synthetically rotating crystal, 
$\boldsymbol{\chi}^{\mathrm{eff}}=
\mathbf{R}\left(\Omega t\right)\cdot\boldsymbol{\chi}_{S}\cdot\mathbf{R}^{-1}\left(\Omega t\right)$. Here, $\boldsymbol{\chi}_{S}$ is a static, non-time-modulated, susceptibility tensor (i.e. the effective crystal being rotated) whose anisotropic properties emerge from the symmetries of the crystal providing the optical Kerr effect. Specifically, for isotropic, trigonal ($3m$, $\overline{3}m$ and $32$) and hexagonal ($622,6mm,6/mmm\,\mathrm{and}\,\bar{6}m2$) crystals, we have 
$\boldsymbol{\chi}_{S}=
\left(3\widehat{\mathbf{x}}\widehat{\mathbf{x}}+\widehat{\mathbf{y}}\widehat{\mathbf{y}}\right)
\chi_{xxxx} E_{0}^{2}$, providing the spatiotemporal permittivity discussed above.
Other crystal symmetries result in different phenomena. For example, hexagonal ($6$, $\overline{6}$ and $6/m$) and trigonal ($3$ and $\overline{3}$) crystals result in the synthetic rotation of an effective anisotropic crystal with nonreciprocal susceptibility 
$\boldsymbol{\chi}_{S}=
\left\{ \left(3\widehat{\mathbf{x}}\widehat{\mathbf{x}}+\widehat{\mathbf{y}}\widehat{\mathbf{y}}\right)\chi_{xxxx}+\left(\widehat{\mathbf{x}}\widehat{\mathbf{y}}-3\,\widehat{\mathbf{y}}\widehat{\mathbf{x}}\right)\chi_{xyyy}\right\} E_{0}^{2}$, further extending the landscape of synthetic rotations (see \cite{SM} for details). In general, our results highlight that crystal symmetries and anisotropic nonlinearities can be harnessed to generalize synthetic motion and spatiotemporal modulation phenomena.
 
A centrosymmetric crystal may also be preferred to avoid competition with second-order nonlinear effects. For example, calcite is a centrosymmetric trigonal crystal \cite{New2011introduction} with a high-frequency cut-off that enables third-harmonic generation into the ultraviolet \cite{Shi2019THG,Penzkofer1988picosecond}. However, the small nonlinear index of Kerr crystals (e.g., $\Delta n \simeq 10^{-7}\,$cm$^2$/GW for calcite \cite{Adair1989nonlinear}) makes it challenging to observe ultrafast phenomena, although frequency-domain effects may still be within reach. In addition, the optical Kerr effect can be strengthened with the use of optical resonators and/or metasurfaces with resonant unit-cells, providing a compromise between modulation strength and speed. 

State-of-the-art experiments on spatiotemporal metamaterials are being driven by doped semiconductors \cite{Zhou2020broadband,Bohn2021spatiotemporal,Tirole2023double,Lustig2023time,Harwood2024super}, which exhibit a much stronger nonlinear response but suffer from slower recovery times and lack polarization-resolved nonlinearities. Nevertheless, the desired anisotropic response could be engineered through appropriate nonlinear metasurface designs \cite{Schirato2020transient,Crotti2024giant} and more sophisticated pumping schemes \cite{Duggan2019optically,Duggan2020nonreciprocal}. Finally, experimental demonstrations at microwave and radio frequencies could be more easily carried out via electronic modulation \cite{Moussa2023observation,Wang2023metasurface,Park2022revealing,Reyes2015observation}.  

{\it Concluding remarks.--}
Our results demonstrate that synthetic crystal rotations mimic the interaction of light with rotating bodies at low rotation frequencies. At higher frequencies, however, synthetic rotations diverge from their mechanical counterparts due to the absence of relativistic effects. This allows access to regimes where the rotation frequency approaches or exceeds the frequency of light, enabling qualitatively new phenomena in the interaction of light with rotating systems. For example, we have demonstrated that the interaction of a short pulse with a synthetically rotating crystal generates spatiotemporal light with nonseparable polarization-pulse shape properties. This advanced structured light could have applications in optical manipulation, such as designing continuously varying optical torque sequences or enabling coherent control to excite multiple polarized transitions within a single pulse. We have also showed that increasing the rotation frequency alters the distribution of frequency--SAM locked sidebands, suggesting the potential generation of frequency-polarization entanglement. In general, we expect our results to stimulate further research into rotation-induced optical phenomena, such as vacuum friction, Casimir forces, and entanglement generation, from the perspective of synthetic motion.

\begin{acknowledgements}
The authors acknowledge support from Grants No.~PID2022-137845NB-C21 and PID2022-137569NB-C42 funded by MICIU/AEI/10.13039/501100011033 and FEDER, EU.
\end{acknowledgements}

\bibliography{library}

\end{document}